\documentclass{article}
\usepackage{color}

\usepackage{amsmath}
\usepackage{amssymb}
\usepackage{cite}

\def\be{\begin{equation}}
\def\ee{\end{equation}}
\def\bea{\begin{eqnarray}}
\def\eea{\end{eqnarray}}

\begin{document}
\renewcommand{\baselinestretch}{1.2}

\noindent{\large\textbf{Phase transition of generalized two
dimensional Yang-Mills U($\textit{N}$) on the sphere for
$G(z)=z^4+\lambda\,z^3$ and Maxwell construction }}\vskip .75 cm

\noindent{\large{~~~~Leila Lavaei}

\small{Department of Physics, Qom University of Technology, Qom,
Iran

E-mail: yalda57L@yahoo.com}}\vskip .75 cm

The large-N behavior of the quartic-cubic generalized two
dimensional Yang-Mills U(N) on the sphere is investigated for
finite cubic couplings. First, it is shown that there are two
phase transitions one of which is third order and the other one is
second order. Second, $gYM_2$ and Maxwell construction are
compared and a relationship between two-dimensional space-time,
that is purely mathematical, and four-dimensional space-time is
obtained.

\section{introduction}

The $YM_2$ theory is defined by the lagrangian $tr(F^2)$ on a
compact Riemann surface, where F is the 2-form field strength. If
one considers $i\,tr(BF)+tr(B^2)$ as the Lagrangian of this
theory, where B is an auxiliary pseudo-scalar field in the adjoint
representation of the gauge group, and uses path-integral method
over the field B, an effective Lagrangian of the form $tr(F^2)$ is
concluded.

Because of two reasons, it is interesting that $YM_2$ theory is
generalized. First, invariance under area-preserving
diffeomorphisms and the lack of propagating degrees of freedom
that are two important properties of $YM_2$ are not unique to the
$i\,tr(BF)+tr(B^2)$ Lagrangian but it is possible to generalize
the theory without losing these two properties. These generalized
theories ($gYM_2$) are defined by replacing the $tr(B^2)$ term by
an arbitrary class function $\textit{f}\,(B)$[5]. Second, it is
conceivable that one of the generalized 2D models will reveal
features which are more relevant and more closely resemble the
four dimensional theories of interest.

Two dimensional Yang-Mills theory ($YM_2$) and generalized
Yang-Mills theories ($gYM_2$s) have been a subject of extensive
study during recent years [1-18]. They are important theories
because they are integrable. It was seen that there are certain
relations between these theories and string theories. These
relations can see from the study of the large $N$ behavior of a
$YM_2$ and $gYM_2$ based on a gauge group $SU(N)$ that is shown in
[3,6,7,9]. On the other hand, these theories can shed light on
some basic features of $QCD_4$.

Because $YM_2$ and $gYM_2$ are integrable models so they are
useful for exploring the general properties of $QCD$. For example,
one can study the large-N behavior of the free energy of this
theories. For this, one must start from the partition function of
one of this theories on a certain surface. Then the sum over
reducible representations of $U(N)$ (or $SU(N)$), appearing in the
expressions of partition function, must be replaced by a path
integral over continuous Young tableaus, and calculated the
area-dependence of the free energy from the saddle-point
configuration. In [20], the logarithmic behavior of the free
energy of $U(N)$ $YM_2$ on a sphere with area $A<A_c=\pi^2$ was
obtained, and in [13] for $A>A_c$ it was proved that there is a
third-order phase transition in $YM_2$. A fact that was known
earlier in the context of lattice formulation [21]. In the case of
$gYM_2$ models, the same transition was shown for $G(z)=z^4$ in
[18] and for $G(z)=z^6$ and
 $G(z)=z^2+\lambda\,z^4$ in [22], and also for
 $G(z)=z^4+\lambda\,z^3$ (for small $\lambda$) in [19], all on the
 sphere.

 In [19], the large $N$ behavior of a $gYM_2$ based on the gauge group
 $U(N)$ on the sphere was studied, for which $G(z)=z^4+\lambda\,z^3$ and
 $\lambda$ was small. It was seen that in that model, the density had two
 maximum with different heights. There was a transition when the absolute
 maximum exceeded one and also there was the other transition when the relative
 maximum exceeded one. So, there were two transitions both
 of which were third order.

 In this paper the large $N$ behavior of a $gYM_2$ based on a gauge group
 $\textit{U}$($\textit{N}$) on a sphere is also studied, for which
 $G(z)=z^4+\lambda\,z^3$ and $\lambda$ is not necessarily small. In Sec.
 2, it is seen that there is a zeroth-order phase transition in free
 energy that is shown in next sections that is not correct.
 In Sec. 3, the zeroth-order phase transition is investigated
 by studying the density and it is obtained that there are four
 segments in each of which the density behaves differently. In
 Sec. 4, the segment 2) is studied and it is resulted that this segment
 must be removed. In Sec. 5, the segment 3) is studied and a
 third-order phase transition in free energy is obtained. In Sec.
 6, the segment 3) is studied again and it is seen that there is
 the other phase transition that is second order. In Sec. 7, Maxwell
 construction is studied. In Sec. 8, $gYM_2$ and Maxwell construction are compared
  and it is seen that the main parameters in
 $gYM_2$ (that are purely mathematical parameters) are similar to
 physical parameters in the four-dimensional space-time. So, it is
 seen that there is a relationship between two-dimensional space-time
 and four-dimensional space-time. Section 9 is
 devoted to the concluding remarks.

\section{The free energy}
 First, it is helpful to review the expression for the partition
 function of a $gYM_2$ on a sphere in the large $N$ limit [18, 19].
 The partition function of the $gYM_2$ on the sphere is [9, 10]
  \bea
 Z=\sum_r d_r^2\,e^{-A\Lambda(r)},
 \eea
where $r$'s are the irreducible representations of the gauge
group, $d_r$ is the dimension of the $r$th representation, $A$ is
the area of the sphere and $\Lambda(r)$ is
 \bea
 \Lambda(r)=\sum_{k=1}^p \frac{a_k}{N^{k-1}}\,C_k(r),
 \eea
in which $C_k$ is the $k$th Casimir of group, and $a_k$'s are
arbitrary constants. If one considers the gauge group $U(N)$ and
parameterizes its representation by $N$ integers $n_1\geq n_2\geq
...\geq n_N$, it is seen that [23]
 \bea
 &&d_r=\prod_{1\leq i\leq j\leq N}
 \left(1+\frac{n_i-n_j}{j-i}\right),\nonumber\\
 &&C_k=\sum_{i=1}^N \left[(n_i+N-i)^k-(N-i)^k\right].
 \eea
For the partition function (1) be convergent, it is necessary that
$p$ in (2) be even and $a_p$ be positive.

In the large $N$ limit, the partition function (1) can be
rewritten as a path integral over continuous parameters. If the
continuous function $\phi(x)$ is introduced as
 \bea
 \phi(x)=-n(x)-1+x,
 \eea
where
 \bea
 0\leq x:=i/N\leq 1,
 \eea
and
 \bea
 n(x):=n_i/N,
 \eea
then the partition function (1) becomes
 \bea
 Z=\int \, \prod_{0\leq x\leq 1}\, d\phi(x)\,e^{S[\phi(x)]},
 \eea
where
 \bea
 S(\phi)=N^2\left[-A\int_0^1 dx\, G[\phi(x)]+\int_0^1\int_0^1
 dy\,log|\phi(x)-\phi(y)|\right],
 \eea
apart from an unimportant constant, and
 \bea
 G(\phi)=\sum_{k=1}^p (-1)^k\, a_k\phi^k.
 \eea
As $N\rightarrow\infty$, for determining the action (8), one
should maximize $S$. The saddle point equation for $S$ is
 \bea
 g[\phi(x)]=P\int_0^1 \frac{dy}{\phi(x)-\phi(y)},
 \eea
where
 \bea
 g(\phi)=\frac{A}{2}\,G'(\phi),
 \eea
and $P$ is the principal value of the integral. If the density is
introduced as
 \bea
 \rho[\phi(x)]=\frac{dx}{d\phi(x)},
 \eea
then (10) becomes
 \bea
 g(z)=P\int_b^a\frac{\rho(\xi)\,d\xi}{z-\xi},
 \eea
and the normalization condition
 \bea
 \int_b^a\rho(\xi)\,d\xi=1.
 \eea
There is also another condition on the density, that is
 \bea
 0\leq\rho(\xi)\leq 1.
 \eea
The above condition can be obtained from the condition $n_1\geq
n_2\geq ...\geq n_N$. To solve (13), the function $H(z)$ is
defined in the complex $z$-plane as [24]
 \bea
 H(z):=\int_b^a\frac{\rho(\xi)\,d\xi}{z-\xi}.
 \eea
One can obtain [18]
 \bea
 H(z)=g(z)-\sqrt{(z-a)(z-b)}\,\sum_{m,n,q=0}^\infty\frac{(2n-1)!!(2q-1)!!}{2^{n+q}n!q!(n+q+m+1)!}\,a^nb^qz^mg^{(n+m+q+1)}(0),\nonumber\\
 \eea
and
 \bea
 \rho(z)=\frac{\sqrt{(a-z)(z-b)}}{\pi}\,\sum_{m,n,q=0}^\infty\frac{(2n-1)!!(2q-1)!!}{2^{n+q}n!q!(n+q+m+1)!}\,a^nb^qz^mg^{(n+m+q+1)}(0),
 \eea
where $g^{(n)}$ is the $n$th derivative of $g$. Using (16) and
(14), it is seen that $H(z)$ behaves like $z^{-1}$ for large $z$.
So, from (17), one can arrive at
 \bea
 &&\sum_{n,q=0}^{\infty}\frac{(2n-1)!!(2q-1)!!}{2^{n+q}n!q!(n+q)!}\,a^nb^qg^{(n+q)}(0)=0,\\
 &&\sum_{n,q=0}^{\infty}\frac{(2n-1)!!(2q-1)!!}{2^{n+q}n!q!(n+q-1)!}\,a^nb^qg^{(n+q-1)}(0)=1.
 \eea
These equations should be used to obtain $a$ and $b$. Defining the
free energy as
 \bea
 F:=-\frac{1}{N^2}\,\ln Z,
 \eea
one can obtain
 \bea
 F'(A)=\int_0^1 dx\,G[\phi(x)]=\int_b^a
 dz\,G(z)\,\rho(z),
 \eea
that $F'(A)$ is the derivative of the free energy with respect to
$A$. Using (9) and (22), it is seen that to obtain $F'(A)$, one
should calculate the integrals
 \bea
 \int_b^a dz\,z^n\,\rho(z).
 \eea
Using (16), by expanding $H(z)$ for $z\rightarrow\infty$, these
integrals appear. So, using (17), one can calculate these
integrals.

If one considers
 \bea
 G(z)=z^4+\lambda z^3,
 \eea
by the rescaling $\tilde z=z\slash\lambda$, one has
 \bea
 G(\tilde z)=\lambda^4\left[\tilde z^4+\tilde z^3\right].
 \eea
Using (12), one can obtain
  \bea
  \tilde \rho(\tilde z)=\lambda\rho(z)
  \eea
where
 \bea
  \tilde\rho(\tilde z):=\frac{dx}{d\tilde z}.
  \eea
Using (16), one has
 \bea
  \tilde H(\tilde z)=\lambda\,H(z)
   \eea
where
 \bea
  \tilde H(\tilde z):=\int_{\tilde b}^{\tilde a} d\tilde\xi\,\frac{\tilde\rho(\tilde\xi)}{\tilde
  z-\tilde\xi}.
 \eea
Using (11) and (17), one obtains
 \bea
  \tilde H(\tilde z)=\tilde A\left[\frac{1}{2}(4\tilde z^3+3\tilde
   z^2)-\frac{1}{4}\sqrt{(\tilde z-\tilde a)(\tilde z-\tilde b)}~\{3(\tilde a+\tilde b+2\tilde z)+3(\tilde a^2+\tilde
  b^2)+8\tilde z^2+2\tilde a\tilde b+4\tilde z(\tilde a+\tilde b)\}\right]
  \eea
where $\tilde A=A\,\lambda^4$. Expanding (30) for large $\tilde
z$, it is seen that
 \bea
  \tilde H(\tilde z)=\alpha_0+\alpha_{-1}\tilde z^{-1}+O(\tilde z^{-2}).
   \eea
As it was mentioned before, $\tilde H(\tilde z)$ should behave
like $\tilde z^{-1}$ for $\tilde z\rightarrow\infty$ . So using
$(30)$, one arrives at two following equations
 \bea
 \tilde\tau^2(3\tilde\sigma+\frac{3}{4})+\tilde\sigma^2(2\tilde\sigma+\frac{3}{2})=0\\
 \frac{3}{4}\tilde\tau^4+\tilde\tau^2(3\tilde\sigma^2+\frac{3}{2}\tilde\sigma)=\frac{1}{\tilde
 A}
 \eea
where
 \bea
  \tilde\sigma:=\frac{\tilde a+\tilde b}{2},\\
  \tilde \tau:=\frac{\tilde a-\tilde b}{2}.
  \eea
Using $(32)$ and $(33)$, one obtains
 \bea
 &&\tilde\tau^2=\tilde\sigma^2\,(\frac{2\tilde\sigma+\frac{3}{2}}{3\tilde\sigma+\frac{3}{4}}),\\
 &&\tilde A=-\frac{9\tilde\sigma^2+\frac{9}{2}\tilde\sigma+\frac{9}{16}}{15\tilde\sigma^6+\frac{45}{2}\tilde\sigma^5
 +\frac{171}{16}\tilde\sigma^4+\frac{27}{16}\tilde\sigma^3}.
 \eea
Using (36) and the condition $\tilde\tau^2\geq 0$, it is concluded
that
 \bea
 -\frac{3}{4}\leq\tilde\sigma\leq-\frac{1}{4}\nonumber.
  \eea
Using (36), it is seen that for $\tilde\sigma=-\frac{1}{4}$,
$\tilde\tau$ is infinity. Also, using (37), it is seen that for
$\tilde\sigma=-\frac{3}{4}$, $\tilde A$ is infinity. So the
condition on $\tilde\sigma$ is converted to
 \bea
 -\frac{3}{4}<\tilde\sigma<-\frac{1}{4}.
  \eea
If one expands (16) for large $\tilde z$ and uses (14), the
derivative of the free energy with respect to the area, using (22)
and (24), becomes
 \bea
  \tilde F'(\tilde A)=\tilde H_4(\tilde z)+\tilde H_5(\tilde z),
   \eea
where $\tilde H_4(\tilde z)$ (or $\tilde H_5(\tilde z)$) is the
coefficient of $\tilde z^{-4}$ (or $\tilde z^{-5}$) in the
expansion of $\tilde H(\tilde z)$. So, expanding $(30)$ for large
$\tilde z$ and using $(37)$, it is seen that
 \bea
  \tilde F'(\tilde A)=-\frac{\tilde\sigma^3(36+390\tilde\sigma+1667\tilde\sigma^2+3500\tilde\sigma^3+3600\tilde\sigma^4+1600\tilde\sigma^5)}
  {12(1+4\tilde\sigma)^2(3+15\tilde\sigma+20\tilde\sigma^2)}.
  \eea
Integrating $(40)$ and using $(37)$, regardless a constant, leads
to
 \bea
  \tilde F(\tilde \sigma)=\frac{1}{4}\big[-\frac{4}{3+4\tilde\sigma}+\frac{3+8\tilde\sigma}{2(3+15\tilde\sigma+20\tilde\sigma^2)^2}
 +\frac{5(1+4\tilde\sigma)}{3+15\tilde\sigma+20\tilde\sigma^2}\nonumber\\
 -4\log|\tilde\sigma|+2\log|1+4\tilde\sigma|-2\log|3+4\tilde\sigma|\,\big].
  \eea
Using (41) and (37), one can plot $\tilde F$ as a function of
$\tilde A$ that is shown in Fig. 2 and also using (40) and (37),
one can plot $\tilde F'$ as a function of $\tilde A$ that is shown
in Fig. 3. Because for each area the free energy must be minimum,
so using Fig. 2, it is concluded that there is a zeroth-order
phase transition in the free energy of the system. Now the
question is wether the zeroth-order phase transition is a correct
result? In the next section, the question will be answered.

\section{The boundary conditions on density}
 Using (37), one can plot $\tilde A$ as a function of $\tilde \sigma$,
 that is shown in Fig. 1. In this figure, it is seen that in the
interval $\tilde A_I\leq \tilde A\leq\tilde A_{II}$, for each
$\tilde A$, there are two or three values for $\tilde\sigma$. So,
using (41), for an area there are several $\tilde F$ in this
interval. Now, one should study the density of the system and
investigate the condition (15). So, using (11) and (18), the
density for $\tilde G(\tilde z)=\tilde z^4+\tilde z^3$ is
 \bea
 \tilde\rho(\tilde z)=\frac{\tilde A}{2\pi}\sqrt{\tilde\tau^2-(\tilde
 z-\tilde\sigma)^2}\,\left[4\tilde\sigma^2+2\tilde\tau^2+4\tilde\sigma\tilde
 z+4\tilde z^2+3(\tilde\sigma+\tilde z)\right].
 \eea
By differentiating $\tilde\rho$ with respect to $\tilde z$ and
putting it equal to zero, one obtains
 \bea
 6(1+4\tilde\sigma)\,\tilde z^3+\{3+6\tilde\sigma(1-4\tilde\sigma)\}\,\tilde
 z^2+\tilde\sigma(-3-6\tilde\sigma+8\tilde\sigma^2)\,\tilde
 z+\tilde\sigma^2(1+2\tilde\sigma)(3+4\tilde\sigma)=0.
  \eea
This equation has three roots for which the density is extremum.
One can plot the roots of the equation and $\tilde b$ and $\tilde
a$ (using (34), (35) and (36)) as a function of $\tilde\sigma$ in
the interval $-\frac{3}{4}<\tilde\sigma< -\frac{1}{4}$. This is
shown in Fig. 4. The density is acceptable only for $\tilde
b\leq\tilde z\leq\tilde a$, because it has been defined just in
the interval. So, using Fig. 4, it is seen that there are three
intervals in each of which the density behaves differently. This
means that one should plot the density in each interval,
separately. These are shown in figures 5,6 and 7. Each of these
figures are related to specific $\tilde\sigma$'s. When
$\tilde\sigma$ is changed, the curves in Fig. 5 and Fig. 6 change
in terms of quantity, but don't change in terms of quality. For
some $\tilde\sigma$'s, Fig. 7 is converted to Fig. 8. One can plot
$\tilde\rho_i$'s as a function of $\tilde\sigma$, that
$\tilde\rho_i$'s are the extremum of $\tilde\rho$ with respect to
$\tilde z$. This is shown in Fig. 9. Using Fig. 9, the boundary of
Fig. 7 and Fig. 8 is obtained. Using Fig. 4 and Fig 9, it is seen
that there are four intervals for $\tilde\sigma$ as follows: 1)
$-\frac{3}{4}<\tilde\sigma\leq\tilde\sigma_I$ that is related to
Fig. 5, 2) $\tilde\sigma_I\leq\tilde\sigma\leq\tilde\sigma_{II}$
that is related to Fig. 6, 3)
$\tilde\sigma_{II}\leq\tilde\sigma\leq\tilde\sigma_{III}$ that is
related to Fig. 7 and 4)
$\tilde\sigma_{III}\leq\tilde\sigma<-\frac{1}{4}$ that is related
to Fig. 8. The condition (using (26) and (15))
 \bea
 0\leq\tilde\rho\leq\lambda
  \eea
restricts the acceptable densities. So, one should redefine the
density in the intervals 2) and 3), because in these intervals the
density is not nonnegative, and then find the free energies of
these redefined densities. So, the free energy of the system is
changed. As a result, the zeroth-order phase transition in the
free energy obtained in previous section is incorrect. In the next
three sections, the density in two intervals 2) and 3) will be
redefined and the order of the phase transition in the free energy
will be obtained.

\section{studying of the second interval}
One can obtain $\tilde\sigma_I=-0.601986$ and
$\tilde\sigma_{II}=-0.418476$, and also $\tilde A_I=37.8042$ and
$\tilde A_{II}=62.2248$. In Fig. 6 it is clear that in this
interval the density is not nonnegative, so it must be redefined.
If one redefines it as
 \bea
 \tilde \rho_2(\tilde z)=\begin{cases} 0,& \tilde z\in[\tilde c,\tilde a]\\
                   \overline\rho_2(\tilde z),& \tilde z \in[\tilde
                   b,\tilde c]
                   \end{cases}
                   \eea
and finds $\overline\rho_2(\tilde z)$ as a function of
$\tilde\delta$ where $\tilde\delta=\frac{\tilde b+\tilde c}{2}$,
it is seen that $\overline\rho_2(\tilde z)$ (as a function of
$\tilde\delta$) is the same as $\tilde\rho_2(\tilde z)$ (as a
function of $\tilde\sigma$). So the behavior of
$\overline\rho_2(\tilde z)$ is similar to that of
$\tilde\rho_2(\tilde z)$ but in the interval $\tilde b\leq\tilde
z\leq\tilde c$. This means that for some $\tilde z$'s,
$\overline\rho_2(\tilde z)$ is negative. So this redefining of the
density is incorrect. If it is redefined as
 \bea
 \tilde \rho_2(\tilde z)=\begin{cases} 0,& \tilde z\in[\tilde d,\tilde e]\\
                   \overline\rho_2(\tilde z),& \tilde z \in[\tilde
                   b,\tilde d]\cup[\tilde e,\tilde a]
                   \end{cases}
                   \eea
one can find [18]
 \bea
    \tilde H_2(\tilde z)=\frac{\tilde A}{2}\left[(4\tilde
    z^3+3\tilde z^2)-\sqrt{(\tilde z-\tilde b)(\tilde z-\tilde
    d)(\tilde z-\tilde e)(\tilde z-\tilde a)}\,\{3+4\tilde
    z+2(\tilde a+\tilde b+\tilde d+\tilde e)\}\right],
      \eea
and
 \bea
  \tilde\rho_2(\tilde z)=\frac{\tilde A}{2\pi}\,\sqrt{(\tilde z-\tilde b)(\tilde z-\tilde
    d)(\tilde z-\tilde e)(\tilde a-\tilde z)}\,\{3+4\tilde
    z+2(\tilde a+\tilde b+\tilde d+\tilde e)\}.
    \eea
Expanding $\tilde H(\tilde z)$ for large $\tilde z$, it is seen
that
 \bea
  \tilde H_2(\tilde z)=\beta_1 \tilde z+\beta_0+\beta_{-1}\tilde
  z^{-1}+O(\tilde z^{-2}).
   \eea
Because $\tilde H_2(\tilde z)$ should behave like $\tilde z^{-1}$,
for large $\tilde z$, so one arrives at three following equations
 \bea
  \beta_1=0,\\
   \beta_0=0,\\
    \beta_{-1}=1.
     \eea
One also has [18]
 \bea
  \int_{\tilde d}^{\tilde e}d\tilde z\{g(\tilde z)-\tilde
  H_2(\tilde z)\}=0,
   \eea
where $g(\tilde z)=\frac{\tilde A}{2}\,G'(\tilde z)$. So one can
obtain
 \bea
  \int_{\tilde d}^{\tilde e}d\tilde z\left[\frac{\tilde A}{2}\sqrt{(\tilde z-\tilde b)(\tilde z-\tilde
    d)(\tilde z-\tilde e)(\tilde z-\tilde a)}\,\{3+4\tilde
    z+2(\tilde a+\tilde b+\tilde d+\tilde e)\}\right]=0.
    \eea
Using this equation and (50), (51) and (52), one can obtain the
four unknowns $\tilde a$, $\tilde b$, $\tilde e$ and $\tilde d$.
To study the structure of the phase transition, one can use the
following change of variables
 \bea
  \tilde b=\tilde b_c(1+P),~~~\tilde d=\tilde d_c(1+M),~~~\tilde e=\tilde
  e_c(1+X),~~~\tilde a=\tilde a_c(1+U),
   \eea
where the index c shows the critical point between intervals 1)
and 2), that is marked with the symbol I in Fig. 1. By
calculation, it is seen that
 \bea
  \tilde b_c=-0.920721,~~~\tilde d_c=\tilde e_c=\tilde
  a_c=-0.283251
   \eea
One can substitute (55) and (56) to (50) and (51), and obtain
 \bea
  P=-0.12M-0.06M^2-0.12U-0.06U^2-0.12X\nonumber\\
 -0.06X^2-0.04MU-0.04UX-0.04MX,\\
  U=-M-0.232M^2-X-0.232X^2-0.232MX.
    \eea
Using these two relations, it can be seen that if $M=-X$ then P,
U, $X^2$ and $M^2$ are of the same order but if $M\neq -X$ then P,
M, X and U are of the same order. If $M=-X$, using (55) and (56),
it is seen that $\tilde a<\tilde e$. This is an incorrect result,
because for $\tilde b$, $\tilde d$, $\tilde e$ and $\tilde a$,
there is the following condition (using (46))
 \bea
  \tilde b<\tilde d<\tilde e<\tilde a,
   \eea
so $M\neq-X$. One can consider
 \bea
  P&=&P_{1/2}\,\Omega^{1/2}+P_1\,\Omega+P_{3/2}\,\Omega^{3/2}+P_2\,\Omega^2,\nonumber\\
  M&=&M_{1/2}\,\Omega^{1/2}+M_1\,\Omega+M_{3/2}\,\Omega^{3/2}+M_2\,\Omega^2,\nonumber\\
  X&=&X_{1/2}\,\Omega^{1/2}+X_1\,\Omega+X_{3/2}\,\Omega^{3/2}+X_2\,\Omega^2,\nonumber\\
  U&=&U_{1/2}\,\Omega^{1/2}+U_1\,\Omega+U_{3/2}\,\Omega^{3/2}+U_2\,\Omega^2,
   \eea
where $\Omega$ is $(\tilde A-\tilde A_c)/\tilde A_c$ and $\tilde
A_c=\tilde A_I$. From (55), (56) and (60), and using (50), (51),
(52) and (54), one can obtain
 \bea
  P&=&-0.067\,\Omega+0.0127\,\Omega^{3/2}+0.05\,\Omega^2,\nonumber\\
  M&=&0.64\,\Omega^{1/2}-0.064\,\Omega-0.266\,\Omega^{3/2}+X_2\,\Omega^2,\nonumber\\
  X&=&0.64\,\Omega^{1/2}-0.064\,\Omega-0.266\,\Omega^{3/2}+X_2\,\Omega^2,\nonumber\\
  U&=&-1.28\,\Omega^{1/2}-0.16\,\Omega+0.34\,\Omega^{3/2}+(0.038-2X_2)\,\Omega^2.
   \eea
Using (61), (55), (56), (48) and $\tilde A=\tilde
A_I\,(1+\Omega)$, one can plot the density as a function of
$\tilde z$ for small $\Omega$. It is seen that for $\Omega\neq 0$,
this graph is similar to Fig. 6. So the redefined density is not
also nonnegative. It follows that the interval 2) must not exist
and the curve between I and II, in figures 1 to 3, must be
removed.

\section{The third-order phase transition}
The interval 3) is $\tilde
\sigma_{II}\leq\tilde\sigma\leq\tilde\sigma_{III}$. One can obtain
$\tilde\sigma_{II}=-0.418476$ and $\tilde\sigma_{III}=-0.33541$,
and also $\tilde A_{II}=62.2248$ and $\tilde A_{III}=25.5689$.
From Fig. 7 it is seen that the density is not nonnegative in this
interval, and so it must be redefined. In line with the previous
section, if one redefines the density as
 \bea
  \tilde \rho_3(\tilde z)=\begin{cases} 0,& \tilde z\in[\tilde d,\tilde e]\\
                   \overline\rho_3(\tilde z),& \tilde z \in[\tilde
                   b,\tilde d]\cup[\tilde e,\tilde a]
                   \end{cases}
                   \eea
it can be concluded that relations (47) to (54) in the interval 2)
also exist in the interval 3). To study the structure of the phase
transition at point $III$, one can use the following change of
variables
 \bea
  \tilde b=\tilde b_{III}(1+P),~~~\tilde d=\tilde d_{III}(1+M),~~~\tilde e=\tilde
  e_{III}(1+X),~~~\tilde a=\tilde a_{III}(1+U),
   \eea
where
 \bea
  \tilde b_{III}=-0.938782,~~~\tilde a_{III}=0.267962,~~~\tilde d_{III}=\tilde
  e_{III}=-0.207295
   \eea
Substituting (63) and (64) for the parameters in (50) and (51), it
is seen that
 \bea
  &&P=-0.047\,M^2-018\,U-0.03\,M X-0.047\,X^2\nonumber\\
  &&U=-0.15\,M^2-0.1\,M X-0.15\,X^2
  \eea
Using these two relations, it is clear that P, U, $M^2$ and $X^2$
are of the same order. So one can consider
 \bea
  P&=&P_1\,\Omega+P_{3/2}\,\Omega^{3/2}+P_2\,\Omega^2,\nonumber\\
  M&=&M_{1/2}\,\Omega^{1/2}+M_1\,\Omega+M_{3/2}\,\Omega^{3/2}+M_2\,\Omega^2,\nonumber\\
  X&=&X_{1/2}\,\Omega^{1/2}+X_1\,\Omega+X_{3/2}\,\Omega^{3/2}+X_2\,\Omega^2,\nonumber\\
  U&=&U_1\,\Omega+U_{3/2}\,\Omega^{3/2}+U_2\,\Omega^2,
   \eea
where $\Omega=\frac{\tilde A-\tilde A_{III}}{\tilde A_{III}}$.
From (63), (64) and (66), and using (50), (51), (52) and (54), it
is obtained
 \bea
  P&=&-0.064\,\Omega+0.048\,\Omega^2,\nonumber\\
  M&=&1.61\,\Omega^{1/2}-0.125\,\Omega-0.498\,\Omega^{3/2}+(-0.01-X_2)\,\Omega^2,\nonumber\\
  X&=&-1.61\,\Omega^{1/2}-0.125\,\Omega+0.498\,\Omega^{3/2}+X_2\,\Omega^2,\nonumber\\
  U&=&-0.535\,\Omega+0.18\,\Omega^2.
   \eea
Using (63), (64), (67), (48) and $\tilde A=\tilde
A_{III}\,(1+\Omega)$, one can plot the density as a function of
$\tilde z$ for small $\Omega$, that is shown in Fig. 10. It is
clear that for $\Omega\neq 0$, the density is nonnegative and so
the redefining (62) in the interval 3) around point $III$ is
correct. Now one can obtain the phase transition in the free
energy around point $III$. Expanding (47) for large $\tilde z$ and
using (39), one can obtain
 \bea
  (\tilde F'_3)_{{}_{III}}(\tilde
  A)=-0.07-0.034\,\Omega+0.0129\,\Omega^2+O(\Omega^3).
  \eea
To calculate the phase transition, one should find $\tilde
F'_w(\tilde A)$ around point $III$. This is the derivative of the
free energy obtained using the initial density (not the redefined
density). Replacing $\tilde\sigma=\Phi+\tilde\sigma_{III}$ in (37)
and using $\tilde A=\tilde A_{III}\,(1+\Omega)$, one can obtain
 \bea
  \Phi=-0.0414\,\Omega+0.00747\,\Omega^2.
  \eea
So using (40), it is obtained
 \bea
  (\tilde F'_w)_{{}_{III}}(\tilde   A)=-0.07-0.034\,\Omega+0.0066\,\Omega^2+O(\Omega^3).
  \eea
Using (68) and (70), one can obtain
 \bea
  (\tilde F'_3)_{{}_{III}}(\tilde A)-(\tilde F'_w)_{{}_{III}}(\tilde A)=0.0063\,\Omega^2+O(\Omega^3).
  \eea
As a result, there is a third-order phase transition in the free
energy around point $III$.

\section{The second-order phase transition}
In Fig. 1 if one moves on the graph from the right side to the
left passing through the point $III$, Fig. 8 is converted to Fig.
7. Redefining the density in the interval 3) similar to section 5
(relation (62)), Fig. 7 will be replaced by Fig. 10. If $\tilde A$
increases continuously, the distance between $\tilde e$ and
$\tilde a$ in Fig. 10 goes to zero. Thus in Fig. 1, the curve
between points $III$ and $II$ will be replaced by the curve
between points $III$ and $IV$ shown in Fig. 11. As a result, in
Fig. 11, the curves between points $III$ and $II$ and between
points $II$ and $I$ must be removed and using the fact that for
each area there is specific density, the curve between $I$ and $V$
must be removed too. This is due to the fact that in the interval
$\tilde\sigma_V<\tilde\sigma<\tilde\sigma_I$, the graph of the
density as a function of $\tilde z$ is similar to Fig. 5 while in
the interval $\tilde\sigma_{IV}<\tilde\sigma<\tilde\sigma_{III}$,
it is similar to Fig. 10. In this section, first, points $IV$ and
$V$ will be found and then the phase transition in the free
energy, to go from $IV$ to $V$, will be obtained.

At point $IV$, $\tilde a$ and $\tilde e$ are equal, because the
density at point $IV$ is the same as one at point $V$. So one can
use the following change of variables
 \bea
  \tilde a=\tilde e=\nu,~~~\tilde d=\gamma+\eta,~~~\tilde  b=\gamma-\eta.
  \eea
Using (50), one obtains
 \bea
  \eta^2=-\frac{4\gamma^2+(\gamma+\nu)(3+4\nu)}{2}.
   \eea
Now, using (51), (52) and the above relation, the following two
equations are obtained
 \bea
 &&-2\gamma\nu(3+4\gamma+4\nu)-\frac{1}{2}(3+8\gamma)\{4\gamma^2+(\gamma+\nu)(3+4\nu)\}=0,\\
 &&-\frac{3}{16}\{\gamma+4\gamma^2-4\gamma\nu-\nu(3+4\nu)\}\{4\gamma^2+(\gamma+\nu)(3+4\nu)\}\tilde  A=1.
   \eea
The roots of the equation (74) are
 \bea
  \nu_1=-\frac{9+48(\gamma+\gamma^2)+\sqrt{81+432\gamma-288\gamma^2-3840\gamma^3-3840\gamma^4}}{24+96\gamma}
  \eea
and
 \bea
  \nu_2=-\frac{9+48(\gamma+\gamma^2)-\sqrt{81+432\gamma-288\gamma^2-3840\gamma^3-3840\gamma^4}}{24+96\gamma}.
  \eea
Using (73) and (54), one can obtain
 \bea
 \int_{\gamma+\eta}^{\nu} d\tilde z\left[\frac{\tilde  A}{2}\,(3+4\tilde z+4\gamma+4\nu)(\tilde z-\nu)\sqrt{\tilde
 z^2-2\tilde  z\gamma+\frac{1}{2}\{6\gamma^2+(\gamma+\nu)(3+4\nu)\}}\right]=0.
  \eea
Now, one should calculate this integral and substitute (76) and
(77) for $\nu$ in the result of the integral, consecutively. Using
$\nu_1$, one can obtain
 \bea
  (\gamma_1=-0.67748,~\tilde A_1=48.2053),~~~(\gamma_2=-0.42,~\tilde A_2=62.2),
  \eea
and using $\nu_2$, one also can obtain
 \bea
 \gamma_3=-0.602,~~~\tilde A_3=37.8042,
 \eea
that (75) has been used to obtain $\tilde A_i$'s. It is seen that
$\nu_1$, $\gamma_2$, and $\tilde A_2$ are related to point $II$
and so $\gamma_2$ is not an acceptable answer, because at this
point the diagram of the density is like Fig. 6 that is not
nonnegative. Also it is seen that $\nu_2$, $\gamma_3$ and $\tilde
A_3$ are related to point $I$. By plotting the density related to
$\tilde A_1$ and $\tilde A_3$ separately, it is seen both of which
are like Fig. 5, and so both of the areas $\tilde A_1$ and $\tilde
A_3$ are apparently correct answers. As a result, one should find
point $IV$ numerically and investigate that wether the correct
answer is $\tilde A_1$ or $\tilde A_3$. Using (50), (51), (52) and
(54) and starting from point $III$, by increasing $\tilde A$
little by little, one can obtain for $\tilde A_I=\tilde A_3$ ,
using the relations in the interval 3), the values $\tilde
b=-0.9069$, $\tilde d=-0.4689$, $\tilde e=-0.09765$ and $\tilde
a=0.2117$. So, one can plot the density related to $\tilde A_I$ as
a function of $\tilde z$. It is seen that the graph is like Fig.
10. So $\tilde A_3$ is not related to point $IV$, because at this
point the graph of the density is like Fig. 5. Increasing $\tilde
A$ further, one can plot the density related to $\tilde A_1$ as a
function of $\tilde z$. It is seen that the diagram is like Fig.
5, and so point $IV$ is specified by $\tilde A_1$, $\gamma_1$, and
$\nu_1$. So the numerical values of the unknown parameters related
to point $IV$ are as follows
 \bea
 \tilde A_{IV}=48.2053,~~~\tilde a_{IV}=\tilde  e_{IV}=0.12023,~~~\tilde d_{IV}=-0.44966,~~~\tilde b_{IV}=-0.905313
 \eea
Using
 \bea
 \tilde b=\tilde b_{IV}(1+P),~~~\tilde d=\tilde  d_{IV}(1+M),~~~\tilde e=\tilde e_{IV}(1+X),~~~\tilde a=\tilde a_{IV}(1+U)
 \eea
and also using (50) and (51), one can obtain
 \bea
 &&P=0.0729\,U+0.0729\,X+0.03699\,U^2+0.03699\,X^2+0.0306\,U X,\nonumber\\
 &&M=-5.584P-0.3257U-0.3257X-0.0938U^2-0.0938X^2-0.0626U X.\nonumber\\
 \eea
Using these two relations, it is clear that if $U=-X$ then $P$,
$M$, $U^2$ and $X^2$ are of the same order, but if $U\neq-X$ then
P, M, U and X are of the same order.

So for $U\neq -X$, one can consider
 \bea
 &&P=P_1\,\Omega+P_2\,\Omega^2\nonumber\\
 &&M=M_1\,\Omega+M_2\,\Omega^2\nonumber\\
 &&U=U_1\,\Omega+U_2\,\Omega^2\nonumber\\
 &&X=X_1\,\Omega+X_2\,\Omega^2,
 \eea
and use (50), (51), (52) and (54), and obtain the unknown
parameters $P_1$ to $X_2$. It is seen that there are two sets of
numerical values for $P_1$ to $X_2$. Using one of these sets, it
is seen that $X=-0.9375\,\Omega$, and $U=-.0057\,\Omega$. Using
(81) and (82), $\tilde a<\tilde e$ is obtained that is incorrect.
Using the other one, it is seen that $\tilde e=\tilde a$, up to
$O(\Omega^2)$. So this is also incorrect. Thus $U\neq -X$ is not
acceptable, and surely $U=-X$.

For $U=-X$, one can expand (50) and (51) up to order P (so up to
order M, $X^2$ and $U^2$) and obtain
 \bea
 P=0.04\,X^2,~~~M=-0.36\,X^2.
 \eea
Using(85), (52) and $\tilde A=\tilde
 A_{IV}\,(1+\Omega)$, it is obtained (up to order $\sqrt\Omega$)
 \bea
 (X)_1=-1.15\,i\sqrt\Omega,~~~(X)_2=1.15\,i\sqrt\Omega.
 \eea
Because $\Omega$ is negative, and $U=-X$, so the correct answer
(up to order $\sqrt\Omega$) is $X=1.15\,i\sqrt\Omega$. Thus one
can obtain
 \bea
 X=1.15\,i\sqrt\Omega+X_1\,\Omega,~~~U=-1.15\,i\sqrt\Omega-X_1\,\Omega,~~~P=-0.057\Omega,~~~M=0.487\Omega.\nonumber\\
 \eea
Expanding $\tilde H(\tilde z)$ for large $\tilde z$, as before,
and then using $\tilde F'_3(\tilde A)=\tilde H_4(\tilde z)+\tilde
H_5(\tilde z)$ around point $IV$, it is seen that
 \bea
 (\tilde F'_3)_{{}_{IV}}(\tilde  A)=-0.09396-0.031046\,\Omega+O(\Omega^2).
 \eea
Now, one should find point $V$ that is in the interval 1) and
$\tilde A_V=\tilde A_{IV}=48.2053$. Because the areas related to
the points $V$ and $IV$ are the same , so the graph of the density
in these two points must be the same in terms of both quality and
quantity. It can be concluded that
 \bea
 \tilde a_V=\tilde d_{IV}=-0.44966,~~~\tilde b_V=\tilde  b_{IV}=-0.905313,~~~\tilde \sigma_V=-0.6775
 \eea
So one can obtain $\tilde F'_w$ around point $V$, like section 5.
Up to order $\Omega$, one can arrive at
 \bea
 (\tilde F'_w)_{{}_{V}}(\tilde  A)=-0.09396-0.013411\,\Omega+O(\Omega^2).
 \eea
Using (88) and (90), it is seen that
 \bea
 (\tilde F'_3)_{{}_{IV}}(\tilde A)-(\tilde  F'_w)_{{}_{V}}(\tilde A)=-0.017635\,\Omega+O(\Omega^2).
 \eea
It is clear that there is a second-order phase transition in the
free energy and $(\tilde F'_3)_{{}_{IV}}(\tilde A)>(\tilde
F'_w)_{{}_{V}}(\tilde A)$, because $\Omega$ is negative. Now, one
can plot $\tilde F'$ as a function of $\tilde A$ that is shown in
Fig. 12. In this figure, the paths from $I$ to $II$, $I$ to $V$,
and $II$ to $III$ are wrong paths and must be removed and the
directed path from $III$ to $IV$ is correct. Meanwhile, in Fig. 12
there are two regions resembling triangles, and the surface area
of these two triangles should be the same. But the proof for that
comes from the fact that one can go from point $III$ to point $IV$
through two different paths, the wrong one which goes through the
lower curve ($III$ to $II$, $II$ to $I$, and then $I$ to $IV$),
and the correct one which goes from $III$ to $IV$ directly. The
difference of the free energies at $IV$ and $III$ should be the
same following both paths. This results that the surface areas of
the triangle-like regions are equal.

\section{Maxwell construction}
 Van der Waals equation of state for a gas is
 \bea
 (p+\frac{an^2}{V^2})(V-nb)=nRT,
 \eea
where $n$ is the number of moles, $T$ is temperature, $p$ is
pressure, and $V$ is the total volume of the gas. $R$ is the gas
constant that is $R=8.3145~\frac{J}{mol.K}$. $a$ and $b$ are
positive experimental constants that are specific for each gas.
This relation, can also be rewritten as
 \bea
 (p+\frac{a'}{v^2})(v-b')=k_BT,
 \eea
where $b'=\frac{b}{N_A}$, $a'=\frac{a}{N_A^2}$, $v=\frac{V}{N}$
($N$ is the total number of particles),
 $k_B=1.38\times 10^{-23}~\frac{J}{K}$ that is Boltzmann's constant, and $N_A=6.02\times 10^{23}$ that is
Avogadro's number. Using above relation, it is seen that in
inflection point, using
 \bea
 \frac{\partial p}{\partial v}|_c=0,~~~~~\frac{\partial^2
 p}{\partial v^2}|_c=0,
 \eea
there are the following three conditions
 \bea
 v_c=3b',~~~T_c=\frac{8a'}{27b'k_B},~~~p_c=\frac{a'}{27b'^2}.
 \eea
The point with these conditions is named critical point. If (93)
is rewritten in terms of $\tilde v$, $\tilde p$ and $\tilde T$
where
 \bea
 \tilde v=\frac{v}{v_c},~~~\tilde p=\frac{p}{p_c},~~~\tilde
 T=\frac{T}{T_c},
 \eea
then (93) is converted to
 \bea
 (3\tilde v-1)(\tilde p+\frac{3}{\tilde v^2})=8\tilde T.
 \eea
It is seen that the above relation is independent of $a'$ and
$b'$. Now, one can plot $\tilde p$ as a function of $\tilde v$
(for van der Waals isotherms), that is shown in Fig.13, for
$T=T_c$, $T>T_c$, and $T<T_c$. By calculating and using the
principle of the least energy, it can be seen that for $T<T_c$,
the curve of isotherm is converted to Fig. 14. In this figure, the
curve between 1 and 2 should be removed and 1 must be connected to
2 directly and also the surface areas B and C should be equal to
each other[25]. If one plots chemical potential as a function of
pressure, that is shown in Fig. 15, compared with Fig. 14, the
region resembling triangle should be removed from the graph. So,
there is a phase transition of first order in chemical potential
in point 1 ( or 2) and as a result there is a phase transition of
second order in free energy of the substance, because chemical
potential is derivative of the free energy.

\section{Comparing $gYM_2$ with Maxwell construction}
Comparing figures 11 and 14 and also figures 12 and 15, it is seen
that $\tilde A$ is equivalent to pressure, $\tilde \sigma$ is
equivalent to volume and $\tilde F'$ is equivalent to chemical
potential. The points $V$ and $IV$ in $gYM_2$ are equivalent to
the points 1 and 2 in Maxwell construction, respectively. To go
from $V$ to $IV$ there is a second-order phase transition in the
free energy of the system, and also the same transition happens to
go from 1 to 2. In Maxwell construction, the surface areas B and C
are equal to each other (in Fig. 14). In $gYM_2$, in Fig. 12, the
surface areas of the triangle-like regions are equal.

If one considers $G(z)=z^4+\lambda z^2$ instead of
$G(z)=z^4+\lambda z^3$, and uses the relations of section 2, it is
obtained
 \bea
 \tilde A=\frac{12}{-1+4\tilde\sigma^4},
 \eea
where $\tilde A=A\,\lambda^2$ and
$\tilde\sigma=\frac{\sigma}{\sqrt{\lambda}}$. Now, $\tilde A$ can
be plotted as a function of $\tilde\sigma$ that is shown in Fig.
16. This figure is similar to Fig. 13 for $T>T_c$.

Also, if one considers $G(z)=z^4+\lambda z$ instead of
$G(z)=z^4+\lambda z^3$, and uses section 2, it is obtained
 \bea
 \tilde A=\frac{48\tilde\sigma^2}{1-16\tilde\sigma^3-80\tilde\sigma^6},
 \eea
where $\tilde A=A\,\lambda^{4/3}$ and
$\tilde\sigma=\frac{\sigma}{\lambda^{1/3}}$. So, one can plot
$\tilde A$ as a function of $\tilde\sigma$ that is shown in Fig.
17. This figure is also similar to Fig. 13 for $T>T_c$.

As a result, in $gYM_2$, the model $G(z)$ plays the role of
temperature.

\section{Concluding remarks}
A $gYM_2$ with quartic and cubic couplings was studied. The effect
of the cubic coupling on the density and the free energy was
investigated. It was seen that there were four intervals that in
two of which the density was not nonnegative and so had to be
redefined. Redefining the density caused two phase transitions in
the free energy, one of which was third order and the other one
was second order. In the end, $gYM_2$ and Maxwell construction
were compared and it was seen that there was a relationship
between two-dimensional space-time, that is purely mathematical,
and four-dimensional space-time.\vspace{.75 cm}

\noindent{\large\textbf{Acknowledgment}}

I would like to thank Mohammad Khorrami for useful discussions. I
would also like to thank the referee for his/her helpful comments
and suggestions to improve this paper.

\newpage
\begin{figure}[ht]
\begin{center}
\begin{picture}(0,0)(0,0)
\includegraphics{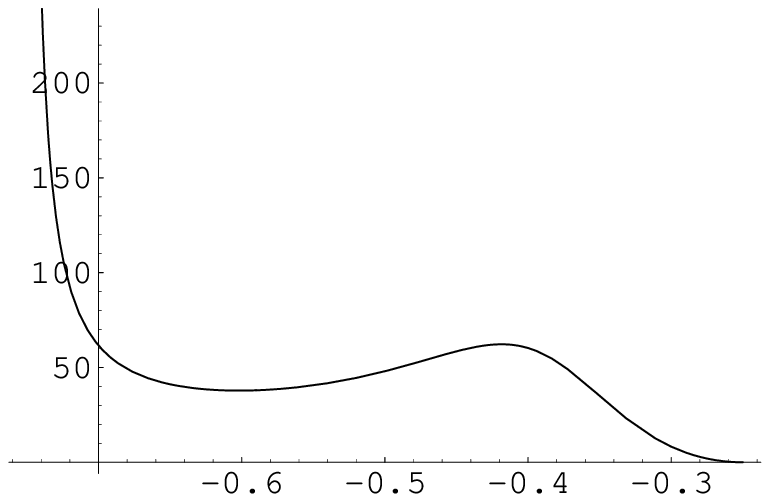}
 \put(-25,-17){$\bullet$} \put(-25,-10){$I$}
 \put(50,-3.5){$\bullet$} \put(48,2.5){$II$}
 \put(84,-25){$\bullet$} \put(86,-18){$III$}
 \put(-70,105){$\tilde A$}
 \put(135,-36){$\tilde\sigma$}
\end{picture}\vspace{1.3cm} \caption{The surface area}
\end{center}
\end{figure}

\begin{figure}[ht]
\begin{center}
\begin{picture}(0,0)(0,0)
\includegraphics{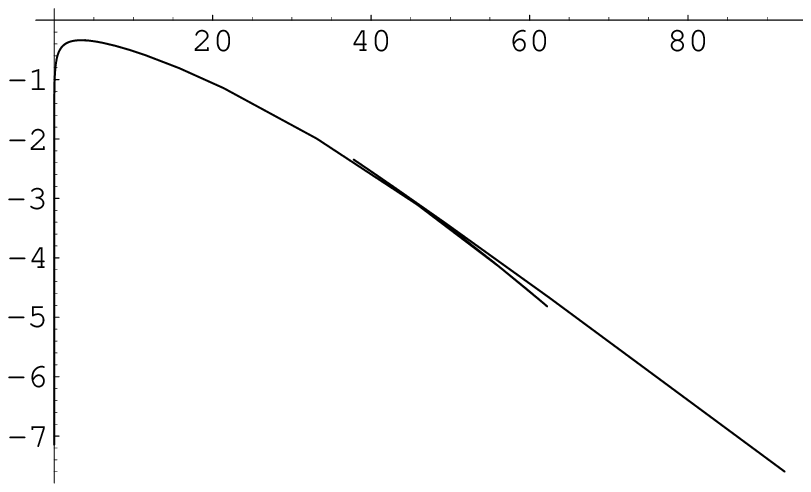}
 \put(-28,-23){$\bullet$} \put(-27,-17){$III$}
 \put(0,-38){$\bullet$} \put(0,-31){$I$}
 \put(55,-80){$\bullet$} \put(50,-88){$II$}
 \put(-88,15){$\tilde F$}
 \put(137,1){$\tilde A$}
\end{picture}\vspace{3.5cm} \caption{The free energy}
\end{center}
\end{figure}

\begin{figure}[ht]
\begin{center}
\begin{picture}(0,0)(0,0)
\includegraphics{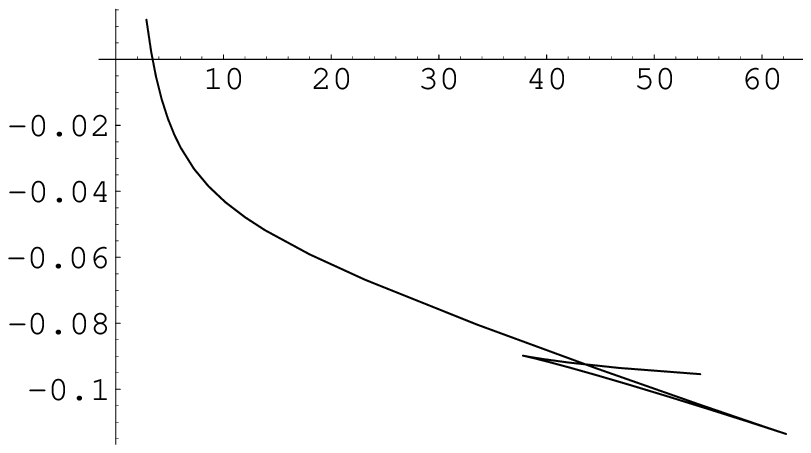}
 \put(13,-101){$\bullet$} \put(12,-94){$III$}
 \put(128,-143){$\bullet$} \put(134,-144){$II$}
 \put(55,-122){$\bullet$} \put(47,-124){$I$}
 \put(-68,-12){$\tilde F'$}
 \put(141,-34){$\tilde A$}
\end{picture}\vspace{4.7cm} \caption{The derivative of the free energy}
\end{center}
\end{figure}

\newpage
\begin{figure}[ht]
\begin{center}
\begin{picture}(0,0)(0,0)
\includegraphics{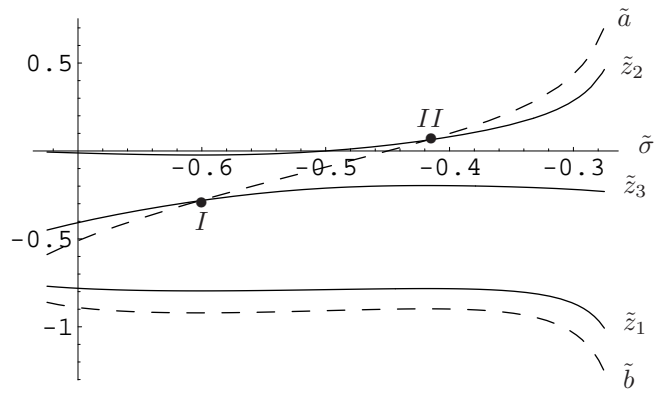}
 \put(-27,41){$\bullet$} \put(-27,33){$I$}
 \put(60,65){$\bullet$} \put(57,72){$II$}
 \put(133,110){$\tilde a$}
 \put(135,-27){$\tilde b$}
 \put(135,-5){$\tilde z_1$}
 \put(134,92){$\tilde z_2$}
 \put(135,47){$\tilde z_3$}
 \put(141,62){$\tilde\sigma$}
\end{picture}\vspace{.5cm} \caption{The roots of the density}
\end{center}
\end{figure}

\begin{figure}[ht]
\begin{center}
\begin{picture}(0,0)(0,0)
\includegraphics{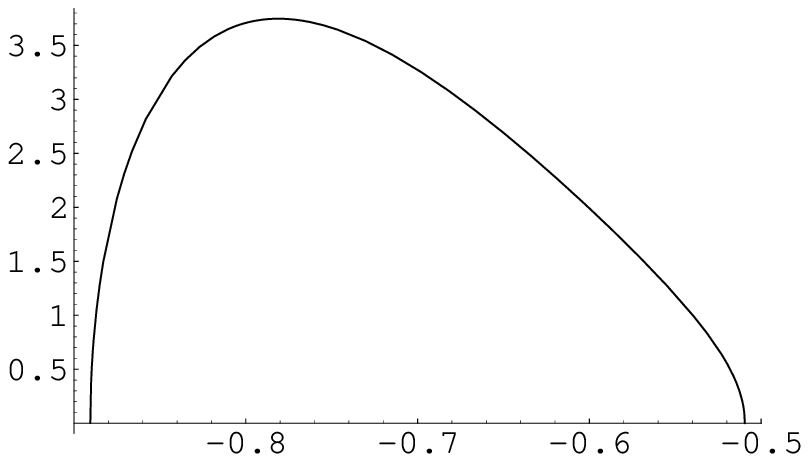}
 \put(-75.5,-107){$\bullet$} \put(-75.5,-120){$\tilde b$}
 \put(114,-107){$\bullet$} \put(118,-99){$\tilde a$}
 \put(-80,22){$\tilde\rho$}
 \put(135,-105){$\tilde z$}
\end{picture}\vspace{3.96cm} \caption{The density in interval 1)}
\end{center}
\end{figure}

\begin{figure}[ht]
\begin{center}
\begin{picture}(0,0)(0,0)
\includegraphics{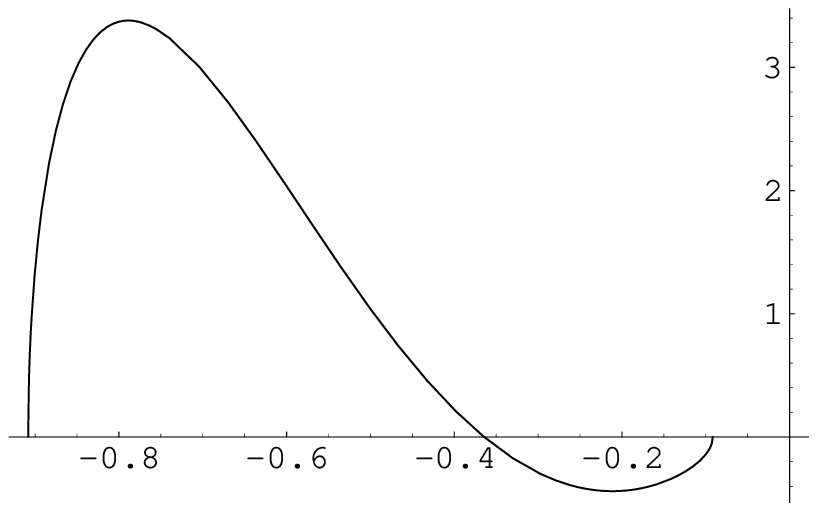}
 \put(-95,-135){$\bullet$} \put(-95,-147){$\tilde b$}
 \put(102.5,-135){$\bullet$} \put(102.5,-128){$\tilde a$}
 \put(123,0){$\tilde\rho$}
 \put(138,-135){$\tilde z$}
\end{picture}\vspace{4.9cm} \caption{The density in interval 2)}
\end{center}
\end{figure}

\newpage
\begin{figure}[ht]
\begin{center}
\begin{picture}(0,0)(0,0)
\includegraphics{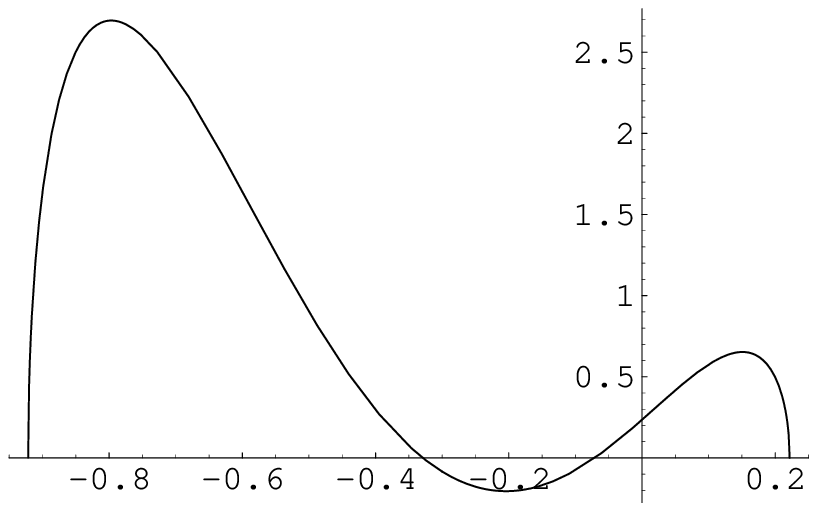}
 \put(-95,-17){$\bullet$} \put(-95,-29){$\tilde b$}
 \put(124,-17){$\bullet$} \put(129,-9){$\tilde a$}
 \put(90,115){$\tilde\rho$}
 \put(140,-15){$\tilde z$}
\end{picture}\vspace{.5cm} \caption{The density in interval 3)}
\end{center}
\end{figure}

\begin{figure}[ht]
\begin{center}
\begin{picture}(0,0)(0,0)
\includegraphics{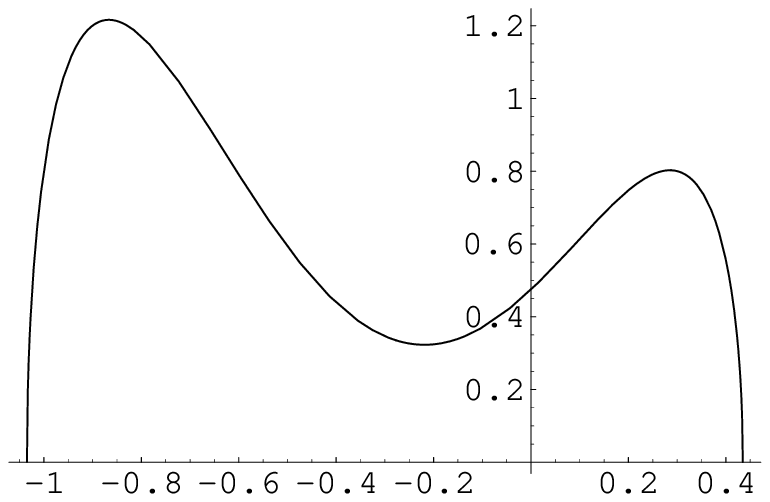}
 \put(-89,-112){$\bullet$} \put(-94,-103){$\tilde b$}
 \put(118,-112){$\bullet$} \put(123,-103){$\tilde a$}
 \put(62,22){$\tilde\rho$}
 \put(140,-110){$\tilde z$}
\end{picture}\vspace{3.96cm} \caption{The density in interval 4)}
\end{center}
\end{figure}

\begin{figure}[ht]
\begin{center}
\begin{picture}(0,0)(0,0)
\includegraphics{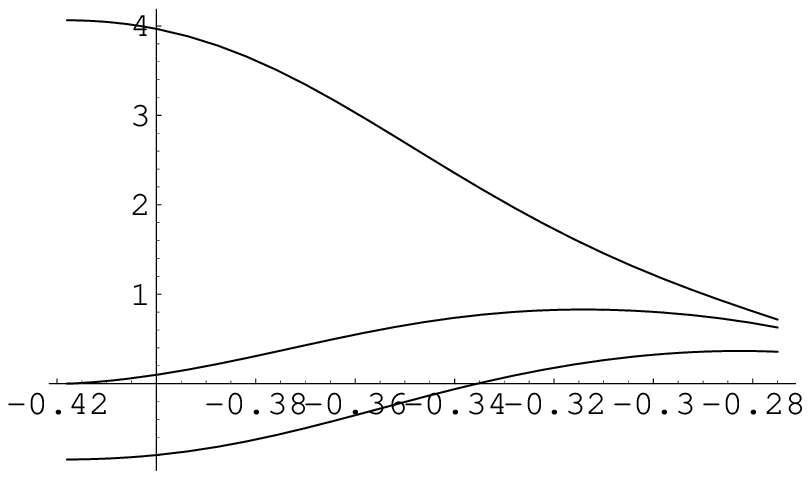}
 \put(-13,-145){$\tilde\rho_3$}
 \put(-18,-103){$\tilde\rho_2$}
 \put(-18,-25){$\tilde\rho_1$}
 \put(-58,-1){$\tilde\rho$}
 \put(140,-123){$\tilde\sigma$}
 \put(36,-125){$\bullet$} \put(33,-118){$III$}
\end{picture}\vspace{4.9cm} \caption{The extremum densities}
\end{center}
\end{figure}

\newpage
\begin{figure}[ht]
\begin{center}
\begin{picture}(0,0)(0,0)
\includegraphics{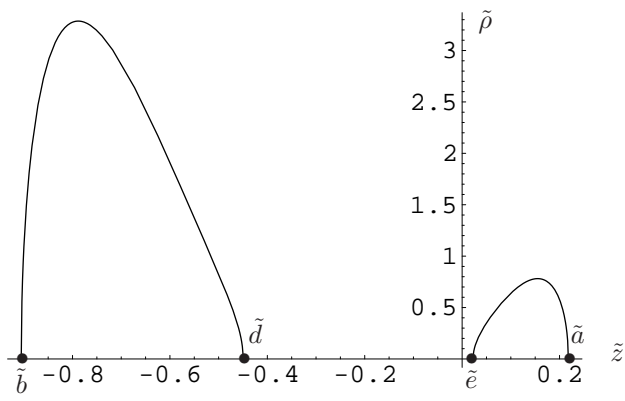}
 \put(-89,-17.5){$\bullet$} \put(-89,-28){$\tilde b$}
 \put(-5,-17.5){$\bullet$} \put(-1,-9){$\tilde d$}
 \put(81,-17.5){$\bullet$} \put(81,-26){$\tilde e$}
 \put(118,-17.5){$\bullet$} \put(121,-9){$\tilde a$}
 \put(86,110){$\tilde\rho$}
 \put(136,-16){$\tilde z$}
 \end{picture}\vspace{.5cm} \caption{The redefined density}
\end{center}
\end{figure}

\begin{figure}[ht]
\begin{center}
\begin{picture}(0,0)(0,0)
\includegraphics{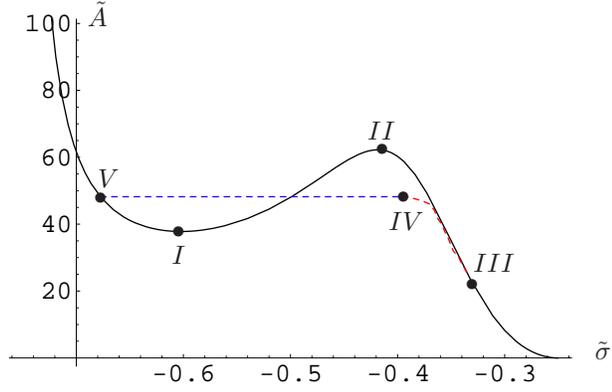}
 \put(-57.5,-52.5){$\bullet$} \put(-56,-46){$V$}
 \put(-28,-65){$\bullet$} \put(-28,-75){$I$}
 \put(49,-34){$\bullet$} \put(46,-28){$II$}
 \put(57,-52){$\bullet$} \put(54,-62){$IV$}
 \put(83,-85){$\bullet$} \put(86,-78){$III$}
 \put(-60,15){$\tilde A$}
 \put(132,-111){$\tilde\sigma$}
\end{picture}\vspace{3.96cm} \caption{The modified surface area}
\end{center}
\end{figure}

\begin{figure}[ht]
\begin{center}
\begin{picture}(0,0)(0,0)
\includegraphics{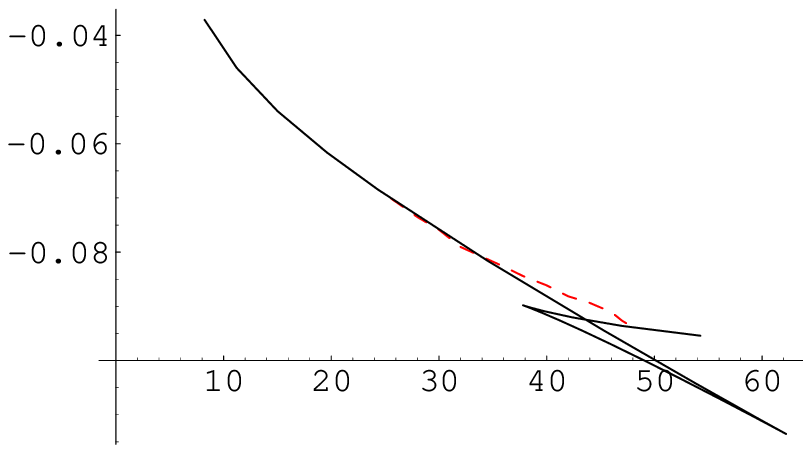}
 \put(10,-75){$\bullet$} \put(9,-68){$III$}
 \put(123,-143){$\bullet$} \put(130,-144){$II$}
 \put(81,-112){$\bullet$} \put(82,-103){$IV,\,V$}
 \put(48,-106){$\bullet$} \put(42,-106){$I$}
 \put(-68,-9){$\tilde F'$}
 \put(139,-121){$\tilde A$}
\end{picture}\vspace{4.9cm} \caption{The modified derivative of the free energy}
\end{center}
\end{figure}

\newpage
\begin{figure}[ht]
\begin{center}
\begin{picture}(0,0)(0,0)
\includegraphics{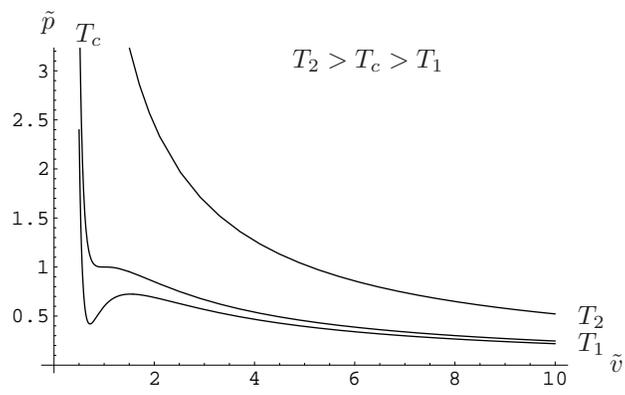}
 \put(-92,105){$T_c$}
 \put(-10,95){$T_2>T_c>T_1$}
 \put(98,-12){$T_1$}
 \put(98,-2){$T_2$}
 \put(-105,110){$\tilde p$}
 \put(110,-20){$\tilde v$}
\end{picture}\vspace{.5cm} \caption{Pressure as a function of volume (isotherms)}
\end{center}
\end{figure}

\begin{figure}[ht]
\begin{center}
\begin{picture}(0,0)(0,0)
\includegraphics{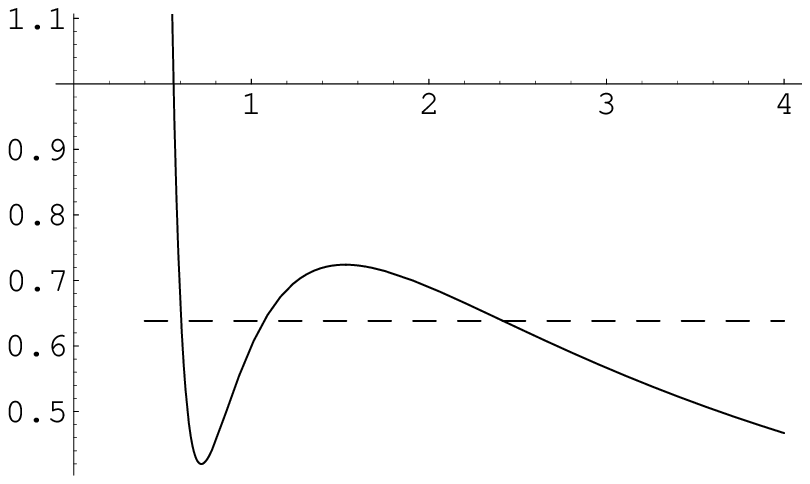}
 \put(-105,-5){$\tilde p$}
 \put(73,-30){$\tilde v$}
 \put(-78,-87){$\bullet$} \put(-83,-94){$1$}
 \put(-3,-87){$\bullet$} \put(-3,-96){$2$}
 \put(-73.5,-118){$\bullet$} \put(-66,-122){$3$}
 \put(-42,-74){$\bullet$} \put(-42,-65){$4$}
 \put(-38,-83){$B$}
 \put(-73,-100){$C$}
\end{picture}\vspace{3.96cm} \caption{Pressure as a function of volume for $T<T_c$}
\end{center}
\end{figure}

\begin{figure}[ht]
\begin{center}
\begin{picture}(0,0)(0,0)
\includegraphics{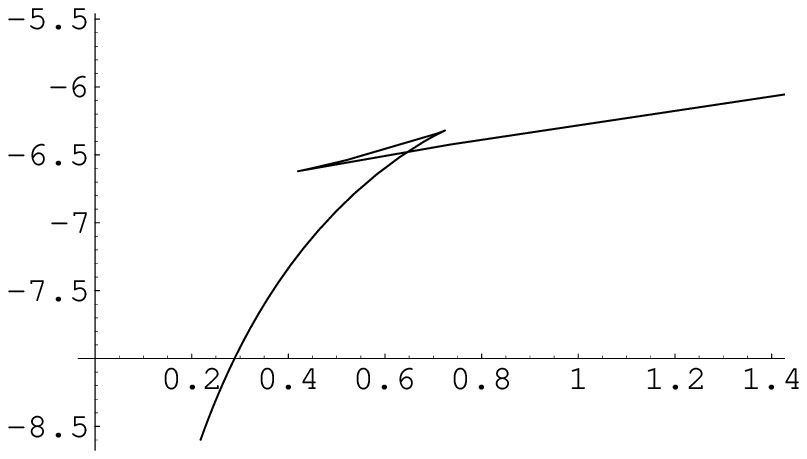}
 \put(-100,-35){$\tilde \mu$}
 \put(73,-125){$\tilde p$}
 \put(-26,-82){$\bullet$} \put(-26,-90){$1$,$2$}
 \put(-52,-85){$\bullet$} \put(-60,-85){$3$}
 \put(-19,-76){$\bullet$} \put(-13,-72){$4$}
 \end{picture}\vspace{4.9cm} \caption{Chemical potential as a function of pressure for $T<T_c$}
\end{center}
\end{figure}

\newpage
\begin{figure}[ht]
\begin{center}
\begin{picture}(0,0)(0,0)
\includegraphics{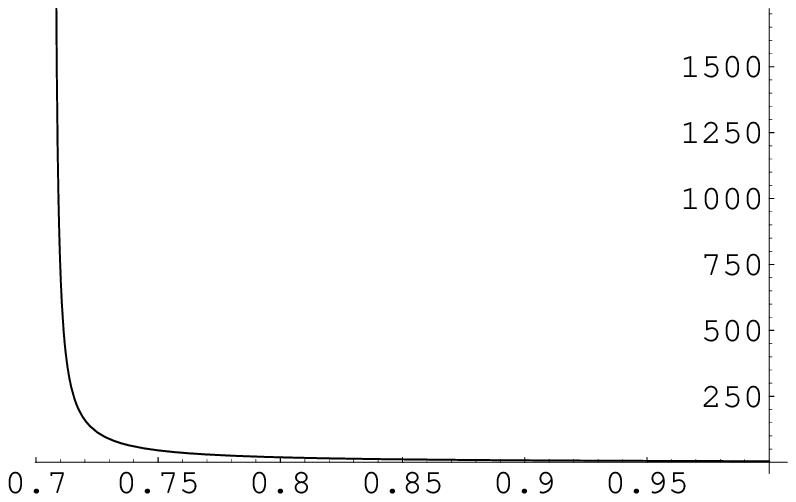}
 \put(90,107){$\tilde A$}
 \put(106,-18){$\tilde \sigma$}
\end{picture}\vspace{.5cm} \caption{Surface area for $G(z)=z^4+\lambda\,z^2$}
\end{center}
\end{figure}

\begin{figure}[ht]
\begin{center}
\begin{picture}(0,0)(0,0)
\includegraphics{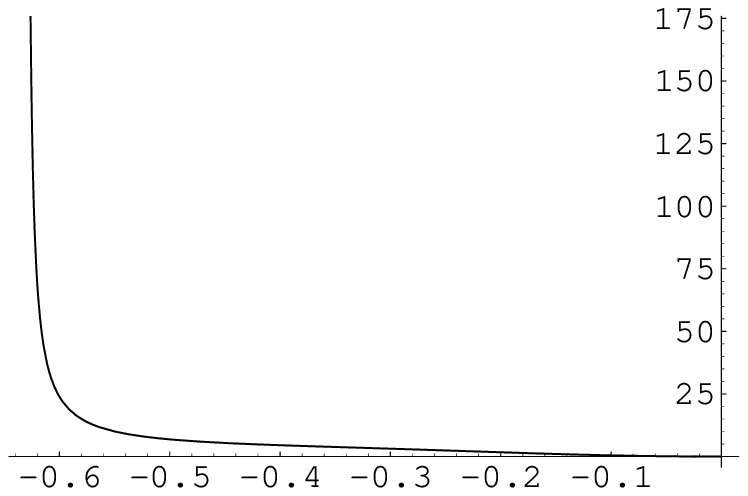}
 \put(85,8){$\tilde A$}
 \put(101,-115){$\tilde \sigma$}
 \end{picture}\vspace{3.96cm} \caption{Surface area for $G(z)=z^4+\lambda\,z$}
\end{center}
\end{figure}


\begin{thebibliography}{99}
\bibitem{EWC} E. Witten, Commun. Math. Phys. 141 (1991) 153.
\bibitem{DFC} D. Fine, Commun. Math. Phys. 134 (1990) 273.
\bibitem{DJG} D.J. Gross, Nucl. Phys. B 400 (1993) 161.
\bibitem{MBGT} M. Blau and G. Thompson, Int. J. Mod. Phys. A 7 (1992) 3781.
\bibitem{EWJG} E. Witten, J. Geom. Phys. 9 (1992)303.
\bibitem{DJWT1} D.J. Gross and W. Taylor, Nucl. Phys. B 400 (1993) 181.
\bibitem{DJWT3} D.J. Gross and W. Taylor, Nucl. Phys. B 400 (1993) 395.
\bibitem{MRDK} M.R. Douglas, K. Lie and M. Staudacher, Nucl. Phys. B 240 (1994) 140.
\bibitem{OGJS} O. Ganor, J. Sonnenschein and S. Yankielowicz, Nucl. Phys. B 434 (1995) 139.
\bibitem{MKMAM} M. Khorrami and M. Alimohammadi, Mod. Phys. Lett. A 12 (1997) 2265.
\bibitem{MAMKI} M. Alimohammadi and M. Khorrami, Int. J. Mod. Phys. A 12 (1997) 1959.
\bibitem{MAMKZ} M. Alimohammadi and M. Khorrami, Z. Phys. C 76 (1997) 729.
\bibitem{MRDV} M.R. Douglas and V.A. Kazakov, Phys. Lett. B 319 (1993) 219.
\bibitem{JAAPP} J.A. Minahan and A.P. Polychronakos, Phys. Lett. B 312 (1993) 155.
\bibitem{JAAPN} J.A. Minahan and A.P. Polychronakos, Nucl. Phys. B 422 (1994) 172.
\bibitem{AAMAMK} A. Aghamohammadi, M. Alimohammadi and M. Khorrami, Mod. Phys. Lett. A 14 (1999) 751.
\bibitem{BRSY} B. Rusakov and S.Yankielowicz, Phys. Lett. B 339 (1994) 258.
\bibitem{MAMKAA} M. Alimahammadi, M. Khorrami and A. Aghamohammadi, Nucl. Phys. B 510 (1998) 313.
\bibitem{L.Lavaei-Yanesi} L.Lavaei-Yanesi and M. Khorrami, Math.
Phys. 49 (2008) 073514.
\bibitem{BR} B. Rusakov, Phys. Lett. B 303 (1993) 95.
\bibitem{DJGE} D.J. Gross, E. Witten. Phys. Rev. D 21 (1980) 446.
\bibitem{MAAT} M. Alimahammadi and A. Tofighi, Eur. Phys. J. C 8
(1999) 711.
\bibitem{VPAP} V.S. Popov and A.M. Perelomov, Sov. Math. Dokl. 8
(1967) 712.
\bibitem{BIPZ} E. Brezin, C. Itzykson, G. Parisi and J.B. Zuber, Commun. Math. Phys.
59 (1978) 35.
\bibitem{KH} Kerson Huang, Statistical mechanics, 2nd edition, chapter 2 (John Wiley and Sons,
1987).
\end{thebibliography}
\end{document}